\documentclass[aps,prl,twocolumn,groupedaddress,showpacs]{revtex4-1}

\usepackage[utf8]{inputenc}
\usepackage[english]{babel}
\usepackage{amsmath,graphicx,enumerate}
\usepackage{epstopdf}

\usepackage{multirow}

\usepackage{color}

\usepackage{bm}
\usepackage[a4paper,includeheadfoot,margin=2.54cm]{geometry}
\usepackage{mathtools}

\DeclareMathAlphabet\mathbfcal{OMS}{cmsy}{b}{n}

\newcommand{\K}{\mathcal{K}}
\newcommand{\KP}{\mathcal{K'}}

\newcommand{\KX}{\mathcal{K}_x}
\newcommand{\KY}{\mathcal{K}_y}
\newcommand{\U}{\mathcal{U}}
\newcommand{\DKX}{\delta k_x}
\newcommand{\DKY}{\delta k_y}
\newcommand{\DU}{\delta u}

\newcommand{\KXP}{\mathcal{K'}_x}
\newcommand{\KYP}{\mathcal{K'}_y}
\newcommand{\UP}{\mathcal{U}'}

\newcommand{\F}{\mathcal{F}}
\newcommand{\G}{\mathcal{G}}
\newcommand{\J}{\mathcal{J}}
 
\newcommand{\DKXEQ}{\delta k_x^{eq}}
\newcommand{\DKYEQ}{\delta k_y^{eq}}
\newcommand{\DUEQ}{\delta u^{eq}}

\newcommand{\DD}{\Delta}
\newcommand{\KS}{\mathcal{K}_s}
\newcommand{\KSP}{\mathcal{K'}_s}

\newcommand{\PA}{{\mathbf{\Pi}}}
\newcommand{\PP}{{\mathbfcal{P}}}
\newcommand{\dP}{\delta \PP}
\newcommand{\B}{\mathbfcal{B}}

\newcommand{\supmat}{Supplementary Information~\cite{sm}}

\begin{document}

\title{Engineered Swift Equilibration of a Brownian Gyrator} 
\author{A. Baldassarri$^{1}$}
\author{A. Puglisi$^{1}$}
\author{L. Sesta$^{2,3}$}
\affiliation{$^1$Istituto dei Sistemi Complessi - CNR and Dipartimento di Fisica, Universit\`a di Roma Sapienza, P.le Aldo Moro 2, 00185, Rome, Italy\\ $^2$Dipartimento di Fisica, Universit\`a di Roma  Sapienza, P.le Aldo Moro 2, 00185, Rome, Italy \\ $^3$Department of Applied Science and Technology (DISAT), Politecnico di Torino, Corso Duca degli Abruzzi 24, Torino, Italy. }

\date{\today}

\begin{abstract}
In the context of stochastic thermodynamics, a minimal model for non equilibrium steady states has been recently proposed: the Brownian Gyrator (BG). It describes the stochastic overdamped motion of a particle in a two dimensional harmonic potential, as in the classic Ornstein-Uhlenbeck process, but considering the simultaneous presence of two independent thermal baths.  When the two baths have different temperatures, the steady BG exhibits a rotating current, a clear signature of non equilibrium dynamics.
Here, we consider a time-dependent potential, and we apply a reverse-engineering approach to derive exactly the required protocol to switch from an initial steady state to
a final steady state in a finite time $\tau$. The protocol can be
built by first choosing an arbitrary quasi-static counterpart - with
few constraints - and then adding a finite-time contribution which
only depends upon the chosen quasi-static form and which is of order
$1/\tau$. We also get a condition for transformations which - in finite time
-  conserve internal energy, useful for applications such as
the design of microscopic thermal engines.
Our study extends finite-time stochastic thermodynamics to
transformations connecting non-equilibrium steady states.

\end{abstract}


\maketitle

{\em Introduction} - Fast switching through two or more modes of
operation in microscopic experiments - where fluctuations dominate -
is a goal for several applications: cyclical mesoscopic thermal
machines such as colloids in time-dependent optical
traps~\cite{blickle2012realization,martinez2015adiabatic,
  schmiedl2007efficiency,bo2013entropic,martinez2016brownian,quinto2014microscopic},
thermal engines realised in bacterial
baths~\cite{krishnamurthy2016micrometre, di2010bacterial}, realisation
of bit operation under noisy environment with connection to
information theory~\cite{berut2012experimental} and much more.
Experiments and theory have recently demonstrated the existence of
special protocols that in finite time realise conditions which are
usually realised in infinite time: these protocols can be deduced by
reverse-engineering the desired, fast, path of evolution of given
observables, including the probability distribution in phase
space~\cite{guery2019shortcuts,guery2014nonequilibrium}.

A paradygmatic example has been given in one effective dimension
with a harmonic trap, realised by optical radiation confining a
colloidal particle~\cite{martinez2016engineered}. The colloidal
particle has reached a steady state in the trap with a stiffness
$k^i$.  Then the trap is modulated from the initial stiffness $k^i$ to
a new stiffness $k^f$ in some finite time $\tau$. If the change $k^i
\to k^f$ is realised in too short a time, e.g. taking ideally $\tau=0$
(what is called ``STEP'' protocol), then the colloidal particle will
take some uncontrolled additional time to reach the steady state
compatible with the final stiffness. Such a ``natural'' time is
related to the typical relaxation times of the system and can be very
long, depending upon the situations. Interestingly, it is possible to
design one or more ``swift equilibration'' (SE) protocols $k(t)$, with
$k(0)=k^i$ and $k(\tau)=k^f$, such that at time $\tau$ the final
steady state is reached and no additional relaxation time is
needed. The shape of $k(t)$ can be non-intuitive: when $\tau$ is
smaller than the typical relaxation times, such protocols may exhibit
large excursions well outside of the range $[k^i,k^f]$. In fact, there
are cases where $k(t)$ can even become negative, posing problems to
its experimental realization. Additional constraints may be introduced
into the mathematical design problem, in order to limit the protocol
excursion~\cite{plata2019optimal}. Other possibilities have been
suggested, where the trap position is also modulated by
additional noise~\cite{chupeau2018thermal}. The SE protocol has been
demonstrated also in an atomic force microscopy
experiments~\cite{le2016fast}.

Here, we discuss the problem of swift equilibration in a
two-dimensional harmonic trap. The generalization may seem a pure
increase of dimensionality, but in fact it allows us to step outside
of the realm of pure equilibrium steady states. A two-dimensional
harmonic trap may be coupled to different thermostats and, in general,
may exhibit rotating currents which break the time-reversal symmetry
even in the steady state~\cite{Ciliberto2013,filliger2007brownian,dotsenko2013two,cerasoli2018asymmetry,villamaina09,argun2017experimental,chiang,soni2017brownian}. 
It is therefore a sound test-ground for the
study of SE protocols for switching between two different
non-equilibrium steady states in a finite time.

{\em Swift equilibration protocol} - Firstly, let us introduce the
general strategy for the SE. We adopted a different and, in a sense,
more general approach with respect to~\cite{martinez2016engineered}.  Consider an
experimental system that can be leveraged controlling some forcing
parameters, which will be noted in a vector $\PA$. The instantaneous
statistical state of the system can be described by a set of
parameters, which will be denoted by a vector $\boldsymbol\gamma(t)$: for
instance in a Gaussian process, as in our case below, these can be the parameters of a
multivariate Gaussian. The value of the parameters $\boldsymbol\gamma(t)$
depends, through a dynamical equation, on the the history of the
applied forcing $\PA$ up to time $t$.  We prepare the system in a
stationary condition, given a value for the forcing $\PA^i$. This
means that we observe a time constant value of the system parameters
$\boldsymbol\gamma$ that depends on the forcing $\PA^i$: we note this value as
$\boldsymbol\gamma^{st}[\PA^i]$.  Our goal is to lead the system into a new
stationary state with a final set of parameters
$\boldsymbol\gamma^{st}[\PA^f]$, {\em in a finite time $\tau$}.

Of course, if $\tau$ is very large, any transformation behaves as a SE
protocol. More precisely, let us consider an arbitrary function
$\PP(s)$, with $s \in [0,1]$, such that $\PP(0)=\PA^i$ and
$\PP(1)=\PA^f$. Then the evolution of parameters $\PA(t)=\PP(t/\tau)$
is a SE protocol in the limit $\tau$ much larger than the
largest characteristic time of the system dynamics. In fact in this
case the change in the forcing parameters is so slow that the system
is always in its stationary state. In this quasi-static forcing,
$\boldsymbol\gamma(t)= \boldsymbol\gamma^{st}[\PP(t/\tau)]$ at any time, including the final
one.

On the contrary, when $\tau$ is finite (smaller than the largest
characteristic time of the system), $\PP(t/\tau)$ is not a SE and must be {\em
  modified} with appropriate finite time corrections, i.e. a
finite-$\tau$ SE protocol reads $\PA=\PP+\frac 1\tau \dP$: here we
denote the finite time corrections with $\dP$.  The quantity $\dP$ ,
hereafter named the finite time correction to the quasi-static
protocol, depends upon the choice of the quasi-static protocol
$\PP$. This is the relevant quantity one has to know to experimentally
perform the desired SE.  The exact, explicit and general formula for $\dP[\PP(s)]$ in the
case of the Brownian Gyrator constitutes the main result of this
letter, and it allows a number of interesting theoretical
considerations.

{\em The Brownian Gyrator} - The system we consider has been
introduced in~\cite{filliger2007brownian} and then studied
in~\cite{dotsenko2013two} and~\cite{cerasoli2018asymmetry}, with an
experimental realization obtained recently in
~\cite{soni2017brownian,argun2017experimental,chiang}. It is widely
known as Brownian Gyrator (BG). Its stochastic differential equation
takes the form
\begin{equation}
\left\{
\begin{array}{rcl}
dx &=& -(k_x \, x + u \, y ) dt + \sqrt{2 T_x} dW_x\\
dy &=& -(k_y \, y + u \, x ) dt + \sqrt{2 T_y} dW_y
\end{array}
\right.\label{FAM}
\end{equation}
which fairly describes an overdamped particle subject to a potential
$V(x,y) = \frac 12 k_x x^2 + \frac 12 k_y y^2 + u x y$ in contact with
two thermal baths at temperature $T_x$ and $T_y$. Note that the
condition of confining potential, required for the steady states, is
$k_x k_y - u^2>0$~\footnote{The concavity condition is necessary to
  reach a steady state but it can be relaxed in transient
  states. However it can become necessary in experimental realizations
  as discussed for instance
  in~\cite{plata2019optimal,chupeau2018engineered,chupeau2018thermal}}. In
  this article we consider the case where $k_x,k_y,u$ may depend upon
  time (on the contrary we keep the temperatures constant). For
  compactness we denote the set of parameters by the vector $\PA$,
  i.e. $\Pi_1=k_x$, $\Pi_2=k_y$ and $\Pi_3=u$.
The associated FP equation reads
\begin{equation}
\partial_t p = \partial_x (p \partial_x V) + \partial_y (p \partial_y V) + T_x \partial_x^2 p +T_y \partial_y^2 p\label{FP}
\end{equation}
where $p(x,y;t)$ is the one time distribution of the stochastic process. The process is Gaussian and for Gaussian initial condition keeps the Gaussian form at all times:
\begin{equation}
  p(x,y;t) =\frac{\exp\left( -\frac 12 \gamma_1 x^2 - \frac 12 \gamma_2 y^2 -\gamma_3\,x\,y\right)}{2\pi\left(\gamma_1 \gamma_2 - \gamma_3^2\right)^{-1/2}},
\label{statdist}
\end{equation}
where $\boldsymbol\gamma = \{\gamma_1,\gamma_2,\gamma_3\}$ depends on time. The introduction of form~\eqref{statdist} in Eq.~\eqref{FP} leads to the equations
governing the time evolution of $\boldsymbol\gamma(t)$, since:
\begin{equation}\label{eqofmotion}
\begin{array}{rcl}
\dot\gamma_1&=& 2 \left(k_x \gamma_1 - T_x \gamma_1^2 + u\gamma_3-T_y\gamma_3^2\right)\\
\dot\gamma_2&=& 2\left( k_y \gamma_2 - T_y \gamma_2^2 + u\gamma_3-T_x\gamma_3^2\right)\\
\dot\gamma_3&=& \gamma_3 (k_x+k_y) + u(\gamma_1+\gamma_2)+\\
& & -2\gamma_3(T_x \gamma_1+T_y\gamma_2).
\end{array}
\end{equation}
If the parameters vector $\PA$ of Eq.~\eqref{FAM} does not depend on
time, then the time-dependence of $\boldsymbol\gamma(t)$ is only due to the
relaxation from initial conditions. In that case, assuming that the
potential is confining, a steady state is reached asymptotically, and -
for ergodicity - coincides with the solution $\partial_t p_{st} =0$,
uniquely determined by the values $\boldsymbol\gamma^{st}[\PA]$ that obey
Eqs.~\eqref{eqofmotion} with all left-hand-sides set to zero (see
\supmat).  When $T_x=T_y=T$ (``thermodynamic equilibrium'') the
Boltzmann distribution $p_{st} \propto e^{-V/T}$ is recovered, i.e.
$\gamma_1^{st} = \frac{k_x}T$, $\gamma_2^{st} = \frac{k_y}T$,
$\gamma_3^{st} = \frac uT$.  On the contrary, when $T_x\neq T_y$, the
steady state is not of the Boltzmann form and, most importantly,
contains a current: ${\bf J}(x,y) = (-p_{st} \partial_x V -
T_x\partial_x p_{st}, -p_{st} \partial_y V-T_y \partial_y p_{st}) \neq
(0,0)$ which is rotational, with null divergence. The steady current
breaks time-reversal invariance (detailed balance) and for this reason
the BG has been proposed as a minimal model for non-equilibrium steady
states~\cite{filliger2007brownian}. 

{\em SE for the Brownian Gyrator} - We look for the forcing protocol
$\PA$ that in a finite time $\tau$ leads the system from the
stationary state $\boldsymbol\gamma^{st}[\PA^i]$ to the stationary state
$\boldsymbol\gamma^{st}[\PA^f]$. We require that the vector $\PA(t) \equiv \{k_x,k_y,u\}$ has the form
$\PA(t)=\PP(t/\tau)+\frac 1\tau\delta \PP(t/\tau)$, where $\PP(s) \equiv \{\KX,\KY,\U\}$ is
a given quasi-static protocol, and $\delta \PP(s)\equiv\{\DKX,\DKY,\DU\}$ is its finite time
correction.

In order to accomplish our task, first we invert the dynamical
equations~(\ref{eqofmotion}) in order to get rid of explicit time and
obtain a set of expressions for $k_x,k_y,u$ as functions of $\gamma_i$
and $\dot{\gamma_i}$ only ($i\in[1,3]$). For the full formula see
\supmat: the important fact is that such expressions can be written
in the form $\PA=\boldsymbol A[\boldsymbol \gamma]+\boldsymbol B[\boldsymbol \gamma] \cdot  \dot{\boldsymbol{\gamma}}$
with $\boldsymbol A[\boldsymbol \gamma]$ a vector and $\boldsymbol B[\boldsymbol \gamma]$ a matrix. If we
require that $\boldsymbol\gamma(t)$ is a function of $s=\frac t\tau$, then $\dot{\boldsymbol
\gamma} = \frac{1}{\tau}\frac{d}{ds}\boldsymbol\gamma$ and the second term
vanishes in the $\tau \to \infty$ limit. Then it is natural to
identify $\PP=\boldsymbol A[\boldsymbol \gamma]$, and $\dP=\boldsymbol B[\boldsymbol \gamma]\cdot
\frac{d}{ds} \boldsymbol\gamma$, see \supmat~for full formula of both terms.

In order to close our loop, now, we need to express everything as a
function of the quasi static protocol. This is done in two steps. The
first step is to invert the relation $\PP=A(\boldsymbol \gamma)$. Since this
relation is valid even in the $\tau\to\infty$ limit, the result is
nothing but the expression of $\boldsymbol\gamma^{st}[\PP]$ that solve
Eqs.~\eqref{eqofmotion} in the stationary condition and parameters set
to $\PP$.  Finally, we have to express $\frac{d}{ds}\boldsymbol\gamma$ as a
function of the quasi-stationary protocols. This is done considering
that $\frac{d}{ds} = \KXP \partial_{\KX}+\KYP
\partial_{\KY}+\UP\partial_{\U}$ and applying this operator to
$\boldsymbol\gamma^{st}[\PP]$ (here and in the following $f'$ stands for $\frac{d}{ds} f$). Putting back $\gamma_i$ and $\frac{d}{ds}\gamma_i$
in the definition of the forcing protocol, we obtain the final
expression:
\begin{equation}
  \begin{array}{rcl}
    \PA(t) &=& \PP(t/\tau) + \frac{1}{\tau} \dP(t/\tau),\\
    \dP(s) &=& \B[\PP(s)] \cdot \PP'(s),
  \end{array}
  \label{protocol2}
\end{equation}
with a matrix $\B$ which is fully defined in the \supmat.
We recall the operative meaning of this formula: one chooses an
arbitrary~\footnote{In the following we always consider continuous functions, which attain the values $\PA^i$ and $\PA^f$ at the border with zero derivatives. However, some of these requirements can be relaxed, if one admits jumps in the finite time forcing $\PA(t)$, as explained in~\cite{plata2019optimal}} quasi-static protocol
$\PP(s)=\{\KX(s),\KY(s),\U(s)\}$ and this corresponds to a
particular form of $\dP(s)=\{\DKX(s),\DKY(s),\DU(s)\}$ for
finite time corrections. 
Before giving a handier expression of $\dP$, we discuss some special cases.

Firstly, we consider the case where there is no interaction among $x$
and $y$, i.e. when $u=0$ both at the beginning and at the end. Then it
does not make sense to switch on $u$ during the protocol, so that the
choice $\U(s)\equiv0$ is quite general (hence $\UP\equiv 0$). The two
degrees of freedom are independent, each one follows a separate
equation and the finite time corrections take the characteristic
log-derivative of the quasi-static protocol: $\DKX =
\frac{1}{2}\frac{\KXP}{\KX}$, $\DKY=\frac{1}{2}\frac{\KYP}{\KY}$, $\DU
=0$. This result coincides with that in~\cite{martinez2016engineered}.

As a second step, we consider the case of two interacting degrees of
freedom $\U\neq 0$ in contact with the same thermal bath at
temperature $T_x=T_y=T$. In this case the finite time corrections to
the quasi-static forcing read:
\[
\begin{array}{rcl}
\DKXEQ &=&  \frac{(\DD+\KY^2) \KXP +  \U^2 \KYP -2 \KY \U  \UP}{2\KS \DD},\\
\DUEQ &=& -\frac{ \U (\KY  \KXP   + \KX \KYP )- 2 \KX  \KY \UP }{2\KS\DD},
\end{array}
\]
where $\KS=\KX+\KY$ and $\DD = \KX \KX - \U^2$. The value for $\DKYEQ$
is obtained swapping the subscripts $x$ and $y$ in the expression for
$\DKXEQ$. Note that the result does not depend on the temperature $T$. 
This result generalizes~\cite{martinez2016engineered} in two dimensions.

A richer phenomenology is obtained in the realm of non-equilibrium,
when $T_x\neq T_y$ and a non zero current appears in the stationary state (a brief study of the current during the SE can be found in~\supmat).

 For instance for $T_y=T_x+dT$, we observe
interesting deviations
\[
\begin{array}{rcl}
\DKX &=& \DKXEQ -\frac{\KS\UP-\U\KSP }{T_x\KS^3}dT+O(dT^2)\\
\DKY &=& \DKYEQ +\frac{\KS\UP-\U\KSP }{T_x\KS^3}dT+O(dT^2)\\
\DU &=& \DUEQ -\frac{\U(\KYP\KX-\KXP\KY)}{T_x\KS^3}dT+O(dT^2).
\end{array}
\]
Note that for a symmetric protocol $\KX=\KY=\K$ the finite time
correction to $u$ becomes of the second order in $dT$.  For such a
symmetric protocol one has that $\DKXEQ=\DKYEQ$ and both are
proportional to a logarithmic derivative, as happens to the non
interacting case: $\DKXEQ = \DKYEQ = \frac 12 \frac{d}{ds} \log
(\K^2-\U^2)$ (which is minus the ``free energy'' of the system divided by $2T$, see \supmat). In the same symmetric case, one can consider a
quasi-static protocol involving only weak interactions $\U \ll 1$. In
this case the general non-equilibrium case reads:
\[
\begin{array}{rcl}
\DKX &=& \frac 12 \frac{\KP}{\K}-\frac {\U}{T_x} \frac{(T_x+T_y)\UP}{4 K^2}+O(\U^2)\\
\DKY &=& \frac 12 \frac{\KP}{\K}-\frac {\U}{T_y} \frac{(T_x+T_y)\UP}{4 K^2}+O(\U^2)\\
\DU &=& \frac 12 \frac{\UP}{\K} -\frac {\U}2 \frac{\KP}{\K^2}+O(\U^2).
\end{array}
\]
Note that the corrections to $\DKX$ and $\DKY$ at first order in $\U$
are different: the finite time correction to a symmetric quasi-static
protocol should not be the same, since the symmetry is broken by the
non equilibrium condition $T_y\neq T_x$. Nevertheless we note that
they differ only by a factor $T_x/T_y$.  It turns out that this is a
general mathematical feature of the solution for the general (non
symmetric) quasi-static protocol. In fact, in the general case, the finite time corrections
$\delta k_{xy}$ and $\delta u$ have a striking mathematical structure:
\begin{equation}\label{final-corrections}
\begin{array}{rcl}
\delta k_{x} &=&\DKXEQ + \frac 1{T_{x}} \frac{d}{ds} \F, \\
\delta k_{x} &=&\DKYEQ - \frac 1{T_{y}} \frac{d}{ds} \F, \\
\delta u &=& \DUEQ + (T_y-T_x)\J + \frac{d}{ds} \G,
\end{array}
\end{equation}
where $\F$ and $\G$ are functions of $\KS$ and $\U$. The
function $\J$ remains finite in the equilibrium limit $T_y\to T_x$,
while $\F$ and $\G$ vanish. The explicit expressions for $\F, \G$ and
$\J$ are quite simple and are given in the \supmat. Equations~\eqref{final-corrections} are the main result of this work.

\begin{figure}[hb]
  \includegraphics[width=7.5cm]{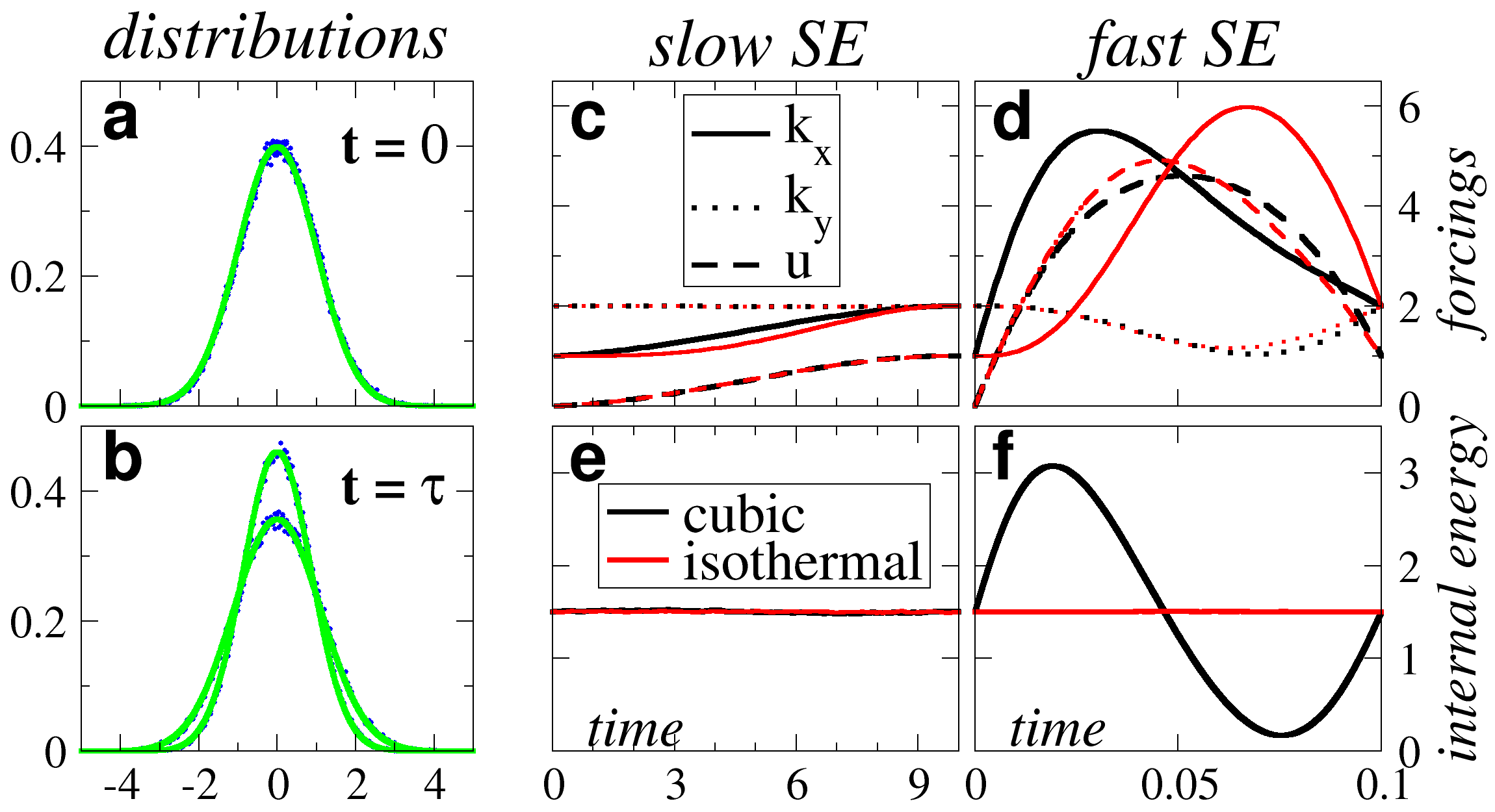}
  \caption{SE in action (details in the \supmat). On the left: a) initial marginal distribution $p(x,t=0)$ and $p(y,t=0)$ (points for simulations) compared with the theoretical stationary distributions (line); b) same quantities at $t=\tau$, the end of the SE transformation. Middle and right plots show the use of  two quasi-static protocols, with the same initial and final parameters $\PA^i$ and $\PA^f$: a simple {\em cubic} protocol (black) and an {\em isothermal} protocol (red). Upper plots: c) forcings for a large $\tau=10$, which are very close to the quasi-static protocol chosen; d) for a much faster SE, $\tau=0.1$, the forcings have large corrections given by Eq.~\eqref{final-corrections}. Lower plots e) and f): the values of the internal energy E in the four different cases ({\em cubic} and {\em isothermal} during slow and fast SE).}\label{figure}
\end{figure}

{\em Energetics} - SE protocols represent an interesting theoretical framework to study energetics and thermodynamics. For instance, we consider internal energy for the model in study:
\[
\begin{array}{rcl}
E&=&\langle V \rangle = \frac{1}{2}k_x\langle x^2\rangle +\frac{1}{2}k_y\langle y^2 \rangle +u\langle x y \rangle,
\end{array}
\]
where, calling $\det\Gamma = \gamma_1 \gamma_2-\gamma_3^2$, we have $\langle x^2
\rangle = \gamma_2 / \det\Gamma$, $\langle
y^2\rangle = \gamma_1/\det\Gamma$ and $\langle
xy \rangle = -\gamma_3/\det\Gamma$.  Since during the SE, the distribution parameters $\boldsymbol\gamma(t)$ depend on the quasi-static protocol as  $\boldsymbol\gamma^{st}[\PP(t/\tau)]$, the expressions for $\langle x^2\rangle$, $\langle y^2\rangle$ and $\langle xy\rangle$ can be written in terms of the quasi-static protocol (see \supmat). 
Hence, using the expression for the
forcing protocols $\PA = \PP + \frac 1\tau \delta \PP$, one can
compute the explicit expression of the internal energy $E$.
Remarkably, it turns out, after careful algebra, that this
expression is simple:
\begin{equation}
E = \frac{T_x+T_y}2 -\frac 1{\tau}  \frac{d}{ds}\left[  \frac{\KX T_y + \KY T_x}{4(\KX \KY - \U^2)}\right]_{s=t/\tau}.
\end{equation}
Several comments about this equation are in order. Firstly we note that
during a quasi-static protocol ($\tau\to\infty$) the internal energy
is constant, as one could expect: this constant value does not depend
on the forcing parameters, neither on $k_{x,y}$ nor on $u$. More
interestingly, the finite time correction to the internal energy,
can be written with a single differential form $\frac 1{\tau} \Theta'$, where $\Theta= \frac{\KX T_y + \KY T_x}{4(\KX \KY - \U^2)}$. An immediate reward of this result is that it
allows to identify a specific class of quasi-static protocols, that we
call the {\it finite time isothermal protocols}, defined as the SE
protocols that keep $\Theta$ constant. Using such a
quasi-static protocol 
one can perform a SE procedure with constant internal energy in any
finite time $\tau$, provided to force the system with
the appropriate finite time corrections~\eqref{final-corrections}. In Fig.~\ref{figure} we show simple numerical simulations giving a demonstration of our results.

{\em Conclusions} - Here we have proposed a general framework for
studying finite time transformations in stochastic processes under the
important request of connecting two steady states without the need of
further relaxation time (``swift equilibration''). Our general
framework is based upon the idea of fixing an arbitrary quasi-static
protocol and then computing the finite-time corrections to it. We have
applied our idea to a model (``Brownian gyrator'') with a harmonic
potential in contact with two different thermal baths, a minimal non-equilibrium generalization of the celebrated Ornstein-Uhlenbeck process. In this sense, 
the model 
can be considered as the harmonic oscillator or the "perfect gas" for non-equilibrium steady
states. Despite the linearity of the model, the problem of SE discloses a rich and promising terrain for theoretical explorations. 
We give the exact explicit expression for the general SE, and also a simple condition to obtain
finite-time transformations that conserve internal energy. The
existence of experimental realizations of the steady Brownian gyrator
~\cite{Ciliberto2013,soni2017brownian,argun2017experimental,chiang} let us foresee interesting
experimental investigations of our procedures in the next future. An
important theoretical perspective concerns the research of
optimal protocols with respect to work or other thermodynamic relevant
quantities, for instance a suitable definition of finite-time
adiabatic transformations~\cite{plata2020finite} to the case where two
thermal baths are present. 

\begin{acknowledgments}
AB and AP acknowledge the financial support of Regione Lazio through the Grant "Progetti Gruppi di Ricerca" N. 85-2017-15257 and from the MIUR PRIN 2017 project 201798CZLJ. LS acknowledges interesting discussions with Andrea Pagnani.
\end{acknowledgments} 

\bibliography{biblio}

\end{document}


\title{Supplementary material for: Engineered Swift Equilibration of a Brownian Gyrator} 
\author{A. Baldassarri$^{1}$}
\author{A. Puglisi$^{1}$}
\author{L. Sesta$^{2,3}$}
\affiliation{$^1$Istituto dei Sistemi Complessi - CNR and Dipartimento di Fisica, Universit\`a di Roma Sapienza, P.le Aldo Moro 2, 00185, Rome, Italy\\ $^2$Dipartimento di Fisica, Universit\`a di Roma  Sapienza, P.le Aldo Moro 2, 00185, Rome, Italy \\ $^3$Department of Applied Science and Technology (DISAT), Politecnico di Torino, Corso Duca degli Abruzzi 24, Torino, Italy.}

\date{\today}

\pacs{} \

\maketitle

\section{SE formulation through quasi-static protocols}

Here we detail the procedure to obtain the expression for the forcing in the SE protocol.

We start from the Fokker-Planck equation of the stochastic process in study:
\begin{equation}
\partial_t p = \partial_x \left[ p\partial_x V + T_x \partial_x p \right]
 + \partial_y \left[ p\partial_y V + T_y \partial_x p \right],\label{FP}
\end{equation}
which can also be written as a continuity equation $\partial_t  p = - \nabla \cdot { \bf J}(p)$ defining a current probability ${\bf J}$, and where the potential $V$ depends on time through a set of forcing parameters $\PA(t) = \{ k_x(t),k_y(t),u(t)\}$ as
\[
 V\left( x,y;\PA(t) \right) = \frac 12 k_x(t)\, x^2 + \frac 12 k_y(t)\, y^2 + u(t)\, x\, y.
 \]
The FP equation is satisfied by the process propagator $p(x,y;t| x_0,y_0;t_0)$, which fully defines our continuous Markov process. However the FP equation is satisfied also by the single time distribution:
\[
p(x,y;t) = \int p(x,y;t| x_0,y_0;t_0) p(x_0,y_0;t_0) dx_0 dy_0
\]
provided the correct initial conditions are imposed
\begin{equation}
\left. p(x,y;t)\right|_{t=t_0} = p(x,y;t_0).\label{FPic}
\end{equation}
In our case, we want that at $t_0=0$ the system is stationary $p(x,y;0) = p_s(x,t)$, where $p_s$ satisfies $\nabla\cdot {\bf  J}(p_s) = 0$, with the forcing parameters in ${\bf J}$ given by the initial values $\PA^i = \{ \KX(0),\KY(0),\U(0)\}$. For the quadratic potential $V$, the stationary solution is Gaussian:
\[
p_s(x,y;\Gamma) = \frac{\sqrt{\gamma^{st}_1 \gamma^{st}_2 - (\gamma^{st}_3)^2}}{2 \pi} \exp\left(-\frac 12 \gamma^{st}_1 x^2 -\frac 12 \gamma^{st}_2 y^2 -\gamma^{st}_3 \,xy\right)
\]
where the vector $\gambold^{st} = \{\gamma_1^{st},\gamma_2^{st},\gamma_3^{st}\}$ depends on the initial value of the forcing $\PA^i$ as:
\begin{eqnarray}
\gamma^{st}_1 &=&  \frac{\left[\KX T_y (\KX+\KY) - \U^2 (T_y-T_x)\right](\KX+\KY)}{T_x T_y (\KX+\KY)^2 + (T_y - T_x)^2 \U^2}\label{gammast1}\\
\gamma^{st}_2 &=&  \frac{\left[\KY T_x (\KX+\KY) + \U^2 (T_y-T_x)\right](\KX+\KY)}{T_x T_y (\KX+\KY)^2 + (T_y - T_x)^2 \ U^2}\label{gammast2}\\
\gamma^{st}_3 &=&  \frac{(\KX T_y + \KY T_x)(\KX+\KY)\U}{T_x T_y (\KX+\KY)^2 + (T_y - T_x)^2 \U^2}.\label{gammast3}
\end{eqnarray}
Note that for $T_x=T_y=T$ the stationary distribution recovers the expected Gibbs distribution $P_s  = Z^{-1} \exp\left(-V/T\right)$, since $\gamma^{st}_1 = \KX/T$, $\gamma^{st}_2 =\KY/T$, and $\gamma^{st}_3 = \U/T$ (and $Z$ is the normalization constant, see later).

Given the initial condition~\eqref{FPic}, with $p(x,y;t_0) = p_s(x,y)$, the solution for a general forcing protocol $\PA(t)$ keeps a Gaussian form, but the parameters $\gambold(t)$ do not satisfy the stationary relation~\eqrefs{gammast1}{gammast3}. In fact, inserting a time dependent Gaussian distribution $P\left(x,y;\gambold(t)\right)$ in \eqref{FP}, it can be easily shown that the FP equation is satisfied as long as:
\begin{eqnarray}
\frac 12 \frac{d\gamma_1}{dt} &=& \gamma_1k_x -T_x \gamma_1^2+u \gamma_3 - T_y \gamma^2_3 \label{gamma1eq}\\
\frac 12 \frac{d\gamma_2}{dt} &=&\gamma_2 k_y -T_y \gamma_2^2+u \gamma_3  - T_x \gamma^2_3 \label{gamma2eq}\\
\frac{d\gamma_3}{dt} &=& \gamma_3 (k_x + k_y) +u (\gamma_1 + \gamma_2)-2 \gamma_3 (\gamma_1 T_x + \gamma_2 T_y).\label{gamma3eq}
\end{eqnarray}
Such equations provide the values of the distribution parameters $\gambold(t) = \{\gamma_1(t),\gamma_2(t),\gamma_3(t)\}$ as a function of the forcings $\PA(t)$. 
Obviously the expressions for $\gambold^{st}$, which do not depend on time, solve these three equations for $k_x=\KX$, $k_y=\KY$, and $u=\U$.

Our SE program is to compute the forcing protocols $\PA(t)$, with the initial values $\PA^i=\PA(0)$ and the final values $\PA^f = \PA(\tau)$ such that $\gambold(\tau) = \gambold^{st}[\PA^f]$ (starting from $\gambold(0) = \gambold^{st}[\PA^i]$).

We look for an expression of the forcing protocol written as the sum of an arbitrary quasi-static protocol plus some finite corrections:  $\PA(t) = \PP(t/\tau)+\frac 1\tau \delta \PP(t/\tau)$. The meaning of the quasi-static protocol $\PP$ is  that in our scheme if we perform the SE in a very large $\tau$, then $\PA \equiv \PP$, or more precisely $\lim_{\tau\to\infty}\PA(s \tau) = \PP(s)$. 

For the Brownian Gyrator we use the following notation:

\begin{eqnarray}
k_x(t) &=& \KX(t/\tau) + \frac 1\tau \delta k_x(t/\tau)\nonumber \\
k_y(t) &=& \KY(t/\tau) + \frac 1\tau \delta k_y(t/\tau)\nonumber \\ 
u(t) &=& \U(t/\tau) + \frac 1\tau \delta u(t/\tau),\nonumber
\end{eqnarray}

that is  $\PA(t) = \{ k_x(t),k_y(t),u(t)\}$,  $\PP(s) = \{\KX(s),\KY(s),\U(s)\}$ and $\delta \PP(s) = \{\DKX(s),\DKY(s),\DU(s)\}$. 

In order to carry on our program, we need to invert the equations~\eqrefs{gamma1eq}{gamma3eq}, so to obtain:

\begin{eqnarray}
k_x &=& \frac{1}{2(\gamma_1+\gamma_2)(\gamma_1\gamma_2 - \gamma_3^2)}\left[ \frac{d\gamma_1}{dt}\gamma_2^2 + \gamma_3^2\left(\frac{d\gamma_2}{dt}-\frac{d\gamma_1}{dt}\right)+ \right.
\left. \frac{d\gamma_1}{dt}\gamma_1\gamma_2-2\gamma_3\gamma_2\frac{d\gamma_3}{dt} \right] +\nonumber\\
& &+ \gamma_1T_x + \frac{\gamma_3^2}{\gamma_1+\gamma_2}\left(T_y-T_x \right)\label{kx} \\
k_y &=& \frac{1}{2(\gamma_1+\gamma_2)(\gamma_1\gamma_2 - \gamma_3^2)} \left[\left(\frac{d\gamma_1}{dt}-\frac{d\gamma_2}{dt}\right)\gamma_3^2+\frac{d\gamma_2}{dt}\gamma_1^2+ \right.
\left. \frac{d\gamma_2}{dt}\gamma_1\gamma_2-2\gamma_1\gamma_3\frac{d\gamma_3}{dt} \right] +\nonumber\\
& &+ \gamma_2T_y + \frac{\gamma_3^2}{\gamma_1+\gamma_2}\left(T_x-T_y\right)\label{ky}\\
u &=& \frac{1}{2(\gamma_1+\gamma_2)(\gamma_1\gamma_2 - \gamma_3^2)}\left[2\frac{d\gamma_3}{dt}\gamma_1\gamma_2 -\gamma_3\frac{d(\gamma_1\gamma_2)}{dt} \right] + \frac{\gamma_3}{\gamma_1+\gamma_2}\left(T_x\gamma_1 + T_y\gamma_2\right).\label{u}
\end{eqnarray}

Now we impose that the distribution parameters depend on time through the rescaled time $s=t/\tau$. As a consequence $\frac{d\gambold}{dt} = \frac 1\tau \gambold'$ (where we indicated with prime the derivative with respect to $s$). Hence, in the large $\tau$ limit $\frac{d\gambold}{dt} \to 0$, and the Eqs.~\eqrefs{kx}{u} become the expressions of the quasi-static protocol as a function of the distribution parameters. They read:

\begin{eqnarray}
\KX = \gamma_1T_x + \frac{\gamma_3^2}{\gamma_1+\gamma_2}(T_y-T_x)\label{KX} \\
\KX = \gamma_1T_x + \frac{\gamma_3^2}{\gamma_1+\gamma_2}(T_y-T_x)\label{KY} \\
\U = \frac{\gamma_3}{\gamma_1+\gamma_2}\left( \gamma_1T_x+\gamma_2T_y \right)\label{U}
\end{eqnarray}

These are nothing but the inversion of the expressions of the stationary parameters $\gambold^{st}$ given in~\eqrefs{gammast1}{gammast3}, where  $\KX$, $\KY$ and $\U$ now depend on (the rescaled) time and we substitute $\gambold^{st}$ with $\gambold(t)$.

If we invert again Eqs.~\eqrefs{KX}{U}, we obtain expressions formally identical to Eqs.~\eqrefs{gammast1}{gammast3}
\begin{eqnarray}
\gamma_1 &=&  \frac{\left[\KX T_y (\KX+\KY) - \U^2 (T_y-T_x)\right](\KX+\KY)}{T_x T_y (\KX+\KY)^2 + (T_y - T_x)^2 \U^2}\label{gamma1}\\
\gamma_2 &=&  \frac{\left[\KY T_x (\KX+\KY) + \U^2 (T_y-T_x)\right](\KX+\KY)}{T_x T_y (\KX+\KY)^2 + (T_y - T_x)^2 \U^2}\label{gamma2}\\
\gamma_3 &=&  \frac{(\KX T_y + \KY T_x)(\KX+\KY)\U}{T_x T_y (\KX+\KY)^2 + (T_y - T_x)^2 \U^2},\label{gamma3}
\end{eqnarray}
but with a completely different meaning: they represent the time dependent distribution parameters $\gambold(t)$ during the SE protocol as a function of the quasi-static protocol $\PP(s=t/\tau)$. We stress that the system in general is not stationary for $0 < t <\tau$, since for a finite time $\tau$, $\PA(t)\neq \PP(s=t/\tau)$. 

The actual forcings $\PA(t)$ are given by the Eqs.~\eqrefs{kx}{u} and we want to express $\PA(t)$ as a function of the quasi-static protocol $\PP$. In order to do this, we have to substitute the expression of $\gambold(t)$ as a function of $\PP$ back in Eqs.~\eqrefs{kx}{u} and we also need the expressions of $\frac{d\gamma}{dt}$ as a function of $\PP$ (and its derivative).
Using the expression of $\gamma$ as a function of $\PP$ in Eqs.~\eqrefs{gamma1}{gamma3},  and  
\[
\frac{d\gambold(t)}{dt} = \frac 1\tau \left[ \KXP \partial_{\KX} + \KYP \partial_{\KY} + \UP\partial_{\U}\right]\gamma
\]
 we can obtain the expressions for $\frac{d\gamma}{dt}$ as a function of $\PP$ and $\PP'$. (As before and in the following the prime denotes the derivatives with respect to $s$, for instance $\PP' \equiv \frac{d \PP}{ds}\left.\right|_{s=t/\tau}$). 
Substituting the expressions found for $\gambold(t)$ (Eqs.~\eqrefs{gamma1}{gamma3}) and $\frac{d\gamma}{dt}$ (not explicitly reported here) as a function of $\PP$ and $\PP'$, back in Eqs.~\eqrefs{kx}{u} we can get the final result in the form $\PA(t) = \PP(s=t/\tau) + \frac 1\tau \delta\PP(s=t/\tau)$, where $\delta\PP(s) = \B[\PP(s)] \PP'(s)$, and $\B$ is a matrix. More explicitly, the expression will be:
\begin{eqnarray}
k_x(t) &=& \left.\left\{\KX(s) + \frac 1\tau\left[ \Bij_{11} \KXP(s) + \Bij_{12} \KYP(s) + \Bij_{13} \UP(s)\right]\right\}\right|_{s=t/\tau}\\
k_y(t) &=&\left. \left\{\KY(s) + \frac 1\tau\left[ \Bij_{21} \KXP(s) + \Bij_{22} \KYP(s) + \Bij_{23} \UP(s)\right]\right\}\right|_{s=t/\tau}\\
u(t) &=& \left. \left\{\U(s) + \frac 1\tau\left[ \Bij_{31} \KXP(s) + \Bij_{32} \KYP(s) + \Bij_{33} \UP(s)\right]\right\}\right|_{s=t/\tau}\\
\end{eqnarray}
The finite time corrections to the quasi static protocols $\delta \PP$, that is  $\delta k_x$, $\delta k_y$ and $\delta u$, are the terms in square brackets involving the elements of the $\B$ matrix. Note that the choice of the quasi-static protocol is completely arbitrary. The only conditions on the arbitrary quasi-static protocol is that it has to satisfy $\PP(0)=\PA^i, \PP(1)=\PA^f$. (Apparently, one should also impose $\PP'(0)=\PP'(1)=0$, but these constraints can be relaxed if one admits jumps in the finite time protocol $\PA(t)$ at the boundaries, as explained in~\cite{plata2019optimal}).

The full expressions for the matrix elements are cumbersome and not really instructive (we give them below, for the sake of completeness). 
However we noted that the expressions for $\delta k_x$, $\delta k_y$ and $\delta u$ can be written in a much more compact and practical form, as we now explain.

Firstly, we consider the equilibrium case $T_x=T_y=T$. In this case, the expressions for $\delta\PP = \delta \PP^{eq}$ are:
\begin{eqnarray}
\delta k_x^{eq} &=& \frac{(\U^2-\KX\KY - \KY^2)\KXP - \U^2 \KYP + 2\KY \U \UP}{2(\KX+\KY)(\U^2-\KX\KY)} \\
\delta k_y^{eq} &=& \frac{(\U^2-\KX\KY - \KX^2)\KYP - \U^2 \KXP + 2\KX \U \UP}{2(\KX+\KY)(\U^2-\KX\KY)}\\
\delta u^{eq} &=& \frac{\KY\U\KXP+\KX\U\KYP-2\KX\KY\UP}{2(\KX+\KY)(\U^2-\KX\KY)}\\
\end{eqnarray}
The interesting observation is that these expressions do not depend explicitly on $T$.  So, even in the general case $T_x\neq T_y$, we can extract the terms composing $\delta \PP^{eq}$ and work with the rest $\delta\PP-\delta\PP^{eq}$ which is the only part explicitly depending on $T_x$ and $T_y$. This corresponds to a (not so simpler) matrix $\dB = \B - \B^{eq}$. However, a careful inspection reveals some striking regularities. For instance it is easy to verify that 
\begin{eqnarray}
\partial_{\KX} \db_{12} &=& \partial_{\KY} \db_{11}\label{schwarzxy}\\
 \partial_{\KX} \db_{13} &=& \partial_{\U}\db_{11} \label{schwarzxu}\\
 \partial_{\KY} \db_{13} &=& \partial_{\U} \db_{12}\label{schwarzyu}.
\end{eqnarray}
This suggests that a single generating function $\F(\KX,\KY,\U)$ can be used to express $\db_{1j}$:
\begin{eqnarray}
\db_{11}&=&\frac 1{T_x} \partial_{\KX} \F\\
\db_{12}&=&\frac 1{T_x} \partial_{\KY} \F\\
\db_{13}&=&\frac 1{T_x} \partial_{\U} \F,
\end{eqnarray}
so that Eqs.~(\eqrefs{schwarzxy}{schwarzyu}) hold for the Schwarz equalities $\partial_\alpha \partial_\beta F = \partial_\beta \partial_\alpha F$ (where $\alpha$ and $\beta$ are two arbitrary arguments of the functions $F$).
The reason for the factor $\frac 1{T_x}$ is that one can show that the symmetry between $\delta k_x$ and $\delta k_y$ (to the exchange of the substripts $x \leftrightarrow y$) implies that:
\begin{eqnarray}
\db_{21}&=&-\frac 1{T_y} \partial_{\KX} \F\\
\db_{22}&=&-\frac 1{T_y} \partial_{\KY} \F\\
\db_{23}&=&-\frac 1{T_y} \partial_{\U} \F.
\end{eqnarray}
The function $\F$ is defined as:
\begin{equation}
\F=\frac 12 \frac{T_x T_y}{T_x-T_y}  \log\left[1+ \frac{\U^2}{(\KX+\KY)^2}  \frac{(T_y-T_x)^2}{T_x T_y}\right] + c. 
\end{equation} 
where $c(T_x,T_y)$ is an arbitrary constant (with respect to $\KX,\KY$ and $\U$).
As a partial final result we have just obtained that:
\begin{eqnarray}
k_{x}(t) &=&\left. \left\{ \KX(s) +\frac 1\tau \left[ \DKXEQ(s) + \frac 1{T_{x}} \frac{d}{ds} \F(\KX,\KY,\U) \right] \right\}\right|_{s=t/\tau},\label{finalkx}\\
k_{y}(t) &=&\left. \left\{ \KY(s) +\frac 1\tau \left[ \DKYEQ(s) - \frac 1{T_{y}} \frac{d}{ds} \F(\KX,\KY,\U) \right] \right\}\right|_{s=t/\tau}. \label{finalky}
\end{eqnarray}
This result is quite useful, even for the numerical computation of the forcing. Once decided the desired (arbitrary) quasi-static protocol, one can compute iteratively the finite time correction with a simple Euler discretization of the derivative $\frac{d\F}{ds}$, without bothering of the cumbersome matrix $\B$.
Unfortunately, a similar trick does not work exactly the same for the three matrix elements $\db_{3j}$ of the matrix $\dB$.
In fact one can show that the "Schwarz" conditions~\eqrefs{schwarzxy}{schwarzyu} do not hold for such elements. However we discovered that if we subtract the coefficient of the first order contribution in $T_y-T_x$, which is
\[
\J = \frac{(\KY \KXP-\KX\KYP) \U}{(T_y \KX+T_x\KY)(\KX+\KY)^2} 
\] 
we can again express the rest of $\dB$ via a generating function $\G$:
\[
\G=\frac{\sqrt{T_x T_y}}{T_y-T_x}
\left\{
		\arctan\left[
			 \frac{T_y-T_x}{\sqrt{T_x T_y}}\cdot \frac{\U}{\KY+\KX}
			 \right]
			 \right\}
	- \frac{\U}{\KY+\KX} + d,
\]
where $d$ is a constant, as $c$.
More precisely, we get that the final expression for the forcing $u$ is:
\begin{equation}
u(t) =  \left. \left\{ \U(s) +\frac 1\tau \left[ \DUEQ(s) + (T_y-T_x)\J(s) + \frac{d}{ds} \G(\KX,\KY,\U)\right] \right\}\right|_{s=t/\tau}.
  \label{finalu}
 \end{equation}
The equations Eqs.~\eqrefs{finalkx}{finalu} are the main result of this work.

\section{The matrix $\B$}
For the sake of completeness we provide the explicit expression of the matrix $\B$, which in its more compact form read:

\small
\begin{eqnarray}
{\cal B}_{11} &=&\frac{{\U }^{2} \left( T_y -T_x  \right) ^{2}}{{T_x } \left( T_y -T_x  \right)\left( \KY +\KX  \right) ^{3}}
 \left( 1+{\frac {{\U }^{2} \left( T_y -T_x  \right) ^{2}}{T_x 
T_y  \left( \KY +\KX  \right) ^{2}}} \right) ^{-1}-{
\frac {\KX \KY +{\KY }^{2}-{\U }^{2}}{2 \left( \KY +\KX 
 \right)  \left( -\KX \KY +{\U }^{2} \right) }}
\\
{\cal B}_{12} &=&\frac{{\U }^{2} \left( T_y -T_x  \right) ^{2}}{{T_x } \left( T_y -T_x  \right)\left( \KY +\KX  \right) ^{3}}
 \left( 1+{\frac {{\U }^{2} \left( T_y -T_x  \right) ^{2}}{T_x 
T_y  \left( \KY +\KX  \right) ^{2}}} \right) ^{-1}-{
\frac {{\U }^{2}}{2 \left( \KY +\KX  \right)  \left( -\KX \KY+{\U }^{2} \right) }}
 \\
{\cal B}_{13} &=& \frac{\U  \left( T_y  -T_x   \right) ^{2}}{{T_x  } \left( -T_y  +T_x   \right) \left( \KY +\KX  \right) ^{2}} \left( 1+{
\frac {{\U }^{2} \left( T_y  -T_x   \right) ^{2}}{T_x  T_y  
 \left( \KY +\KX  \right) ^{2}}} \right) ^{-1}+{\frac {\KY \U }
{ \left( \KY +\KX  \right)  \left( -\KX \KY +{\U }^{2}
 \right) }}
\\
{\cal B}_{21} &=&\frac{ {\U }^{2} \left( T_y -T_x  \right) ^{2}}{{T_y } \left( T_x-T_y\right)  \left( \KY +\KX  \right) ^{3}} \left( 
1+{\frac {{\U }^{2} \left( T_y -T_x  \right) ^{2}}{T_x T_y 
 \left( \KY +\KX  \right) ^{2}}} \right) ^{-1}-{\frac {{\U }^
{2}}{2 \left( \KY +\KX  \right)  \left( -\KX \KY +{\U }^{2}
 \right) }}
\\
{\cal B}_{22} &=&\frac{{\U }^{2} \left( T_y -T_x  \right) ^{2}}{{T_y } \left( T_x -T_y  \right)\left( \KX +\KY  \right) ^{3}}
 \left( 1+{\frac {{\U }^{2} \left( T_y -T_x  \right) ^{2}}{T_y 
T_x  \left( \KX +\KY  \right) ^{2}}} \right) ^{-1}-{
\frac {\KX \KY +{\KX }^{2}-{\U }^{2}}{2 \left( \KY +\KX 
 \right)  \left( -\KX \KY +{\U }^{2} \right) }} \\
{\cal B}_{23} &=& \frac{\U  \left( T_y  -T_x   \right) ^{2}}{{T_y  } \left( T_y  -T_x   \right) \left( \KY +\KX  \right) ^{2}} \left( 1+{
\frac {{\U }^{2} \left( T_y  -T_x   \right) ^{2}}{T_x  T_y  
 \left( \KY +\KX  \right) ^{2}}} \right) ^{-1}+{\frac {\KX \U }
{ \left( \KY +\KX  \right)  \left( -\KX \KY +{\U }^{2}
 \right) }}
\\
{\cal B}_{31} &=& {\frac { \left( T_y  -T_x   \right) \KY \U }{ \left( \KX T_y  +\KY T_x   \right)  \left( \KY +\KX  \right) ^{2}}}-\frac{\U}{ 
 \left( \KY +\KX  \right) ^{2}}\left[\left( 1+{\frac {{\U }^{2} \left( 
T_y  -T_x   \right) ^{2}}{T_x  T_y   \left( \KY +\KX 
 \right) ^{2}}} \right) ^{-1}-1\right]+\nonumber\\
 & &+{\frac {\KY \U }{ 2\left( \KY +\KX 
 \right)  \left( -\KX \KY +{\U }^{2} \right) }}
\\
{\cal B}_{32} &=& -{\frac { \left( T_y  -T_x   \right) \KX \U }{ \left( \KX T_y  +\KY T_x   \right)  \left( \KY +\KX  \right) ^{2}}}-\frac{\U}{ 
 \left( \KY +\KX  \right) ^{2}}\left[\left( 1+{\frac {{\U }^{2} \left( 
T_y  -T_x   \right) ^{2}}{T_x  T_y   \left( \KY +\KX 
 \right) ^{2}}} \right) ^{-1}-1\right]+\nonumber\\
 & &+{\frac {\KX \U }{ 2\left( \KY +\KX 
 \right)  \left( -\KX \KY +{\U }^{2} \right) }}
\\
{\cal B}_{33} &=&\left( \KY +\KX  \right) ^{-1} \left[\left( 1+{\frac {{\U }^{2} \left( 
T_y  -T_x   \right) ^{2}}{T_x  T_y   \left( \KY +\KX 
 \right) ^{2}}} \right) ^{-1}-1\right] -{
\frac {\KX \KY }{ \left( \KY +\KX  \right)  \left( -K_{{1}
}\KY +{\U }^{2} \right) }}.
\end{eqnarray}

\section{Internal Energy and isothermal protocols}

We consider the internal energy of the system, defined as:

\begin{equation}
E=\int\, V(x,y)\, p(x,y,t)\,dx \,dy = \frac{1}{2}k_x\Sigma_{11}+\frac{1}{2}k_y\Sigma_{22}+u\Sigma_{12},
\end{equation}

where $\Sigma$ is the covariances matrix of the Gaussian propagator $P$,
i.e. $\Sigma_{11}=\langle x^2 \rangle=\gamma_2/\det\Gamma$,
$\Sigma_{22}=\langle y^2\rangle=\gamma_1/\det\Gamma$ and
$\Sigma_{12}=\Sigma_{21}=\langle xy \rangle=-\gamma_3/\det\Gamma$, and $\det \Gamma = \gamma_1 \gamma_2 - \gamma_3^2$. 

During the SE, as for the $\gambold(t)$, the expressions for $\Sigma$ can be written in terms of the quasi-static protocol, and read:

\begin{eqnarray}
\Sigma_{11} &=&  \frac  1{\KX\KY-\U^2} \left[\KY T_x  + \frac{\U^2}{\KX  + \KY }\,(T_y-T_x)\right] = \frac 1{\DELTA} \left[\KY T_x +\frac {\U^2}{\KS} (T_y-T_x)\right] \label{Sigma11}\\
\Sigma_{22} &=&  \frac  1{\KX\KY-\U^2} \left[\KX T_y  - \frac{\U^2}{\KX  + \KY }\,(T_y-T_x)\right]= \frac 1{\DELTA} \left[\KX T_y -\frac {\U^2}{\KS} (T_y-T_x)\right]\label{Sigma22}\\
\Sigma_{12}=\Sigma_{21} &=& -\frac{\U}{\KX \KY - \U^2}\frac {\KX T_y+\KY T_x}{\KX+\KY} = -\frac{\U}{\KS} \frac{\CHI}{\DELTA},\label{Sigma12}
\end{eqnarray}
where $\CHI = \KX T_y + \KY T_x$, $\DELTA=\KX\KY-\U^2$, and $\KS=\KX+\KY$.

Using these expressions together with~\eqrefs{finalkx}{finalu} for the forcing protocols $\PA = \PP + \frac 1\tau \delta \PP$, one can compute the explicit expression of the internal energy $E$. 

Strikingly, it turns out, after a careful algebra, that this expression is very simple:
\begin{equation}
E = \frac{T_x+T_y}2 +\frac 1{\tau}  \frac{d}{ds}\left[ \frac{\KX T_x + \KY T_y}{4(\KX \KY - \U^2)}\right]_{s=t/\tau}.
\end{equation}
Interestingly, the finite time correction to the internal energy, which has to be in the form $E_x \KXP+E_y \KXP + E_u \UP$, with three (cumbersome) coefficients depending on the quasi-static protocol,  can actually be written with a single differential form, since $E_x =  \partial_{\KX} \Theta, E_y= \partial_{\KY} \Theta, E_u = \partial_{\U} \Theta$, where $\Theta \equiv  \frac{\CHI}{4\Delta}$. Note that $2 \Theta'/\tau$ looks like an extra temperature of the system during the forcing.

Choosing a quasi-static protocol satisfying
\begin{equation}
\frac{\KX T_y + \KY T_x}{\KX \KY - \U^2} = \text{constant},\label{isothermal}
\end{equation}
one can perform a SE procedure with constant internal energy in a finite time $\tau$, provided to correct the quasi-static protocol with its appropriate finite time corrections~\eqrefs{finalkx}{finalu}.

\begin{figure}[ht]
  \includegraphics[width=12.0cm]{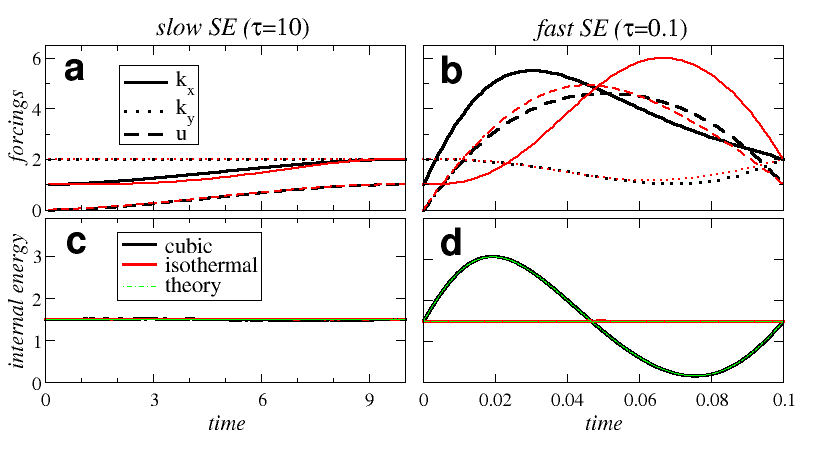}
  \caption{In the upper panels (a) and (b) we show the forcing for the SE used in simulations. In (a) the forcings obtained with the quasi-static protocols defined in~Eqs.~\eqref{cubic} (black), and in Eqs.~\eqref{isotherm} (red) applied during a long time $\tau=10$. The SE is so slow that the forcings $k_x$ (continuous line), $k_y$  (dotted line), and $u$ (dashed line) are very close to the quasi-static protocols. In (b) the same quasi-static protocols are used, but for a much shorter SE, with $\tau=0.1$. Note the large finite time corrections for $k_x$ and $u$, as well as the perturbation of $k_y$, which was almost constant in the slow SE. An other interesting observation, in this case, is that, despite the quasi-static {\em isothermal} protocol~\eqref{isothermal} is very close to the {\em cubic} quasi-static protocol~\eqref{cubic}, their corrections are very different, especially for $k_x$. In the lower panes, the internal energy measured during the different SE are shown. In (c) the internal energy measured during the slow SE ($\tau=10$): the internal energy is almost constant, exactly as for a quasi-static SE, for both the protocols. In (d) the internal energy during the fast SE protocols are shown. The internal energy during the fast cubic SE shows large variations, in contrast with the isothermal protocol, where the internal energy is constant. \label{fig:isothermal}  }
\end{figure}

\begin{figure}[ht]
  \includegraphics[width=12.0cm]{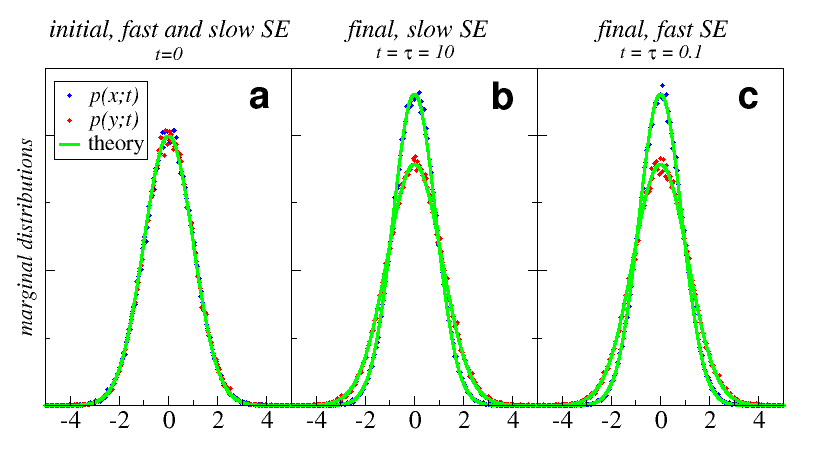}
  \caption{In these figures we show the marginal distribution $p(x;t)=\int p(x,y;t) dy$ (blue points) and $p(y;t)=\int p(x,y;t) dx$ (red points) at the beginning and the end of the SE protocols considered in this work. In (a) we show the initial marginal distributions for all the SE protocols considered: note that since we choose $u(0)=0$, the marginal distributions are of the Gibbs-Boltzmann form $p(x;0) \propto \exp(-k_x x^2/T_x)$ and $p(y;0) \propto \exp(-k_y y^2/T_y)$ (green lines). Note that, despite we consider different temperatures $T_x=1$ and $T_y=2$,  $p(x;t)$ is equal to $p(y;t)$ despite the two degrees of freedom are different: this is a special artifact due to the choice we made for $k_x(0)=1$ and $k_y(0)=2$, so we have $k_x(0)/T_x = k_y(0)/T_y$.  In (b) and (c) we show the marginal distributions at the end of the SE protocols: in this case since $u(\tau)=1$, the marginal distributions have no more the Gibbs-Boltzmann form. The marginal distributions $p(x;\tau)$ and $p(y;\tau)$ are different, but are independent on $\tau$.
  In both cases, $\tau=10$ and $\tau=0.1$, the marginal distribution coincide with the theoretical prediction (green line) that does not depend on $\tau$.
   \label{fig:checkdist} }
\end{figure}

\section{Numerical simulations}
We performed some numerical simulations in order to further check the (exact) relation \eqref{isothermal}.
We considered two different quasi-static protocols. In the first one the parameters, which will be indicated as ``cubic'' in the following, are defined by:
\begin{equation}
\begin{array}{rcl}
\KX(s) &=& 1+3 s^2 - 2 s^3 \\
\KY(s) &=& 2 \\
\U(s) &=& 3  s^2 - 2 s^3 
\end{array}\label{cubic}
\end{equation}

and the initial values of the forcing are $\PA^i=\{1,2,0\}$, while the final values are $\PA^f=\{2,2,1\}$.
The second protocol has the same expression for $\KY$ and $\U$, while $\KX$ is chosen according to Eq.~\ref{isothermal}. The constant is chosen in order to have the same final value of $\KX(1)=2$. The final expressions for this {\em isothermal} protocol become: 
\begin{equation}
\begin{array}{rcl}
\KX(s) &=& \frac{2\left[ (3 s^2-2 s^3)^2 (T_x+T_y)+ 3\, T_x\right]}{4 T_x+ T_y}\\
\KY(s) &=& 2\\
\U(s) &=& 3  s^2 - 2 s^3 
\end{array}
\label{isotherm}
\end{equation}
The initial values for the isothermal protocols are $\PA^i = \{ 6 T_x / (4 T_x + T_y)   ,2,0\}$, while the final values are, as before, $\PA^f=\{2,2,1\}$. In order to have exactly the same initial and final values for the two protocols, we choose $T_x=1$ and $T_y=2$.
In Figure~\ref{fig:isothermal}, in the two upper plots, we show the forcing corresponding to the {\em cubic} and {\em isothermal} protocol for two different values of the total time $\tau$. On the left we show a slow protocol $\tau=10$, where the finite time corrections are very small and the protocol has the same shape of the quasi-static choice. On the right we show a protocol two order of magnitude faster: $\tau=0.1$. In all the plots, the {\em cubic} protocol is in black, while the {\em isothermal} protocol is in red.
Firstly, we also performed a check on the initial and final distributions, as shown in Fig.~\ref{fig:checkdist}. The marginal distributions $p(x;t)$ and $p(y,t)$ are computed for the cubic protocol. We collected the statistics on $10^5$ trajectories for each SE protocol considered. In the figure we show that the final (marginal) distributions are the same irrespectively of the protocol duration $\tau$ and, since the very slow $\tau=10$ case is quasi-static in practice, the final distributions are identical to the stationary distributions. Similar results are obtained for the {\em isothermal} protocol.

Now we consider the internal energy, as shown in the lower plots of Fig.~\ref{fig:isothermal}. In the lower left plot, the slow protocols are shown. As expected, since the finite time corrections to the protocols are negligible, the energy is constant for both the {\em cubic} and the {\em isothermal} protocol. On the other hand, for the fast protocols, while the internal energy shows a large variation during the SE, the isothermal protocol keeps the energy constant, as predicted by our computations. The theoretical predictions for the internal energy can not be distinguished, in the plots, by the simulation results.

\section{Work and "Free energy" for stationary equilibrium transformations}

In a transformation one can always define the {\em average} thermodynamic work~\cite{jarzynski2007comparison} as
\begin{equation}
  W(t) =\int_0^t dt' \langle \partial_t' V(x,y,t') \rangle= \\
  \int_0^t \left[\frac{1}{2} \dot k_x \left\langle x^2\right\rangle+\frac{1}{2} \dot k_y\left\langle y^2\right\rangle + \dot u\left\langle xy\right\rangle\right] dt'.
\end{equation}
Since in our settings we have
$\PA(t)=\PP(t/\tau)+\frac{1}{\tau}\dP(t/\tau)$, hence $\dot \PA = \frac 1\tau \PP' + \frac 1{\tau^2}\dP'$, and  $\Sigma(t) = \Sigma[\PP(t/\tau)]$,
we have
\[
W(t) = \int_0^{t/\tau} w_0(s) ds +  \int_0^{t/\tau} w_1(s) ds
\]
where
\begin{eqnarray}
w_0 &=& \frac 12 \KXP \Sigma_{11} +\frac 12 \KYP \Sigma_{22} + \UP \Sigma_{12}\\
w_1 &=& \frac 1\tau \left[\frac 12 \delta k_x' \Sigma_{11} +\frac 12 \delta k'_y \Sigma_{22} +\delta u' \Sigma_{12}\right]
\end{eqnarray}

Note that the first integral is the work needed/produced during a stationary quasi-static protocol, while the second the correction for a finite time SE protocol.

Using \eqrefs{Sigma11}{Sigma12} one can write the stationary contribution to the work as:
\[
w_0 = \frac{T_x (\KX+\KY)\KY +(T_y-T_x)\U^2}{2(\KY+\KX)(\KX\KY-\U^2)}\KXP
+ \frac{T_y (\KX+\KY)\KX -(T_y-T_x)\U^2}{2(\KY+\KX)(\KX\KY-\U^2)}\KYP
-\frac{\U(T_y \KX+T_x \KY)}{(\KY+\KX)(\KX\KY-\U^2)}\UP
\]

The stationary contribution to the work, for an equilibrium $T_x=T_y=T$ system, can be written as:
\begin{equation}
w_0^{eq} (s) =  \frac{d}{ds} T \log{\sqrt{\KX\KY-\U^2}}
\end{equation}
This equation agrees with equilibrium statistical mechanics. In fact, at equilibrium, we can compute
the partition function
\[
Z = \int dx dy\, e^{-\frac{V(x,y)}T}
\]
which because of the gaussian form of the potential $V(x,y) = \frac 12 k_x x^2 + \frac 12 k_y y^2 + u \,xy$ reads
\[
Z = \frac{2 \pi T}{ \sqrt{k_x k_y-u^2}}
\]
During a stationary SE protocol, one has $k_x=\KX, k_y=\KY$, and $u=\U$, this leads finally to the free energy during the quasi-stationary equilibrium transformation: 
\[
F  = -T \log Z = T \log \sqrt{\KX \KY-\U^2} -T \log{2\pi T}
\]
This allows to rewrite the equilibrium stationary contribution to the work as:
\[
w_0(s) =\frac{d}{ds} F(\KX,\KY,\U) 
\]

If we consider the case of a non interacting protocol:

\[
w_0 = \frac 12 T_x \frac{\KXP}{\KX}+\frac 12  T_y \frac{\KYP}{\KY} 
\]
we note that we can rewrite it in terms of a time derivative:
\[
w_0 = T_x \frac{d}{ds} \log \sqrt{\KX} + T_y  \frac{d}{ds} \log \sqrt{\KY}
\]
This can be easily understood, reasoning as in the previous paragraph, in terms of independent equilibrium of the two degree of freedoms, each one with its own free energy:
\[
F_{x,y} = T_{x,y}\log\sqrt{\KXY} -T_{x,y}\log\sqrt{2\pi T_{x,y}}
\]

\section{Probability current}

A relevant quantity for the Brownian Gyrator is the probability current. In particular, for $T_x\neq T_y$ there can be a stationary non zero current, since the system is out-of-equilibrium. 

It is convenient to consider the {\em mean velocity}~\cite{seifertrev}:
\[
\boldsymbol \nu(x,y,t) = \frac{{\bf J}(x,y,t)}{p(x,y;t)}
\]
During a SE this quantity depends on the quasi-static protocols and its derivative. In the {\em stationary state} (or {\em during an quasi-static SE}) it reads:

\begin{eqnarray}
 \nu_x^{st}(x,y)&=& 
\frac{(T_y - T_x)\, \U  \left\{\left[ \U^2 (T_y-T_x)+\KY T_x  (\KX+\KY)\right] y + \left[\,\U (\KX T_y+\KY  T_x)\right] x \right\}}{\U^2 (T_y-T_x)^2+T_y T_x (\KX+\KY)^2}\\
\nu_y^{st}(x,y)&=& \frac{(T_y- T_x)\, \U  \left\{ \left[\U^2 (T_y-T_x)-\KX T_y (\KX+\KY)\right] x - \left[\,\U (\KX  T_y+\KY T_x)\right] y \right\}}{\U^2 (T_y-T_x)^2+T_y T_x (\KX+\KY)^2},
\end{eqnarray}
which is zero if $T_y=T_x$ or $\U=0$.

During the SE this quantity has finite time corrections proportional to $\KXP,\KYP,\UP$ (not shown here). However, it turns out that these corrections do not contribute to the curl of the mean velocity, which reads {\em during a finite time SE}:
\[
\nabla \times \boldsymbol \nu(x,y,t) = -\frac{(T_y-T_x) (\KX+\KY) (\KX T_y + \KY T_x)\,\U}{\U^2 (T_y-T_x)^2+T_y T_x (\KX+\KY)^2}.
\]
Note that this quantity is homogeneous in space and depends on the quasi-stationary protocol only (hence it depends on time through $t/\tau$).
Again this quantity is zero for $T_y=T_x$ or $\U=0$.

The divergence of the mean velocity is zero in the stationary state, while has finite and homogeneous in space finite time corrections (not shown).

\bibliography{biblio}